%% file: main.tex
\documentclass[conference]{IEEEtran}
\IEEEoverridecommandlockouts
\usepackage{array}
\usepackage[table]{xcolor}
\usepackage{xspace}
\usepackage{amsmath}
\usepackage{circledsteps}
\usepackage{listings}
\usepackage{enumitem}
\usepackage{multirow}
\usepackage{siunitx}
\usepackage{subcaption}
\usepackage[hidelinks]{hyperref}

\usepackage{adjustbox}
\usepackage{makecell}
\usepackage{xurl}
\usepackage{tabularx}
\usepackage{booktabs}
\usepackage{colortbl} 
\newcolumntype{Y}{>{\centering\arraybackslash}X}
\usepackage{caption}
\usepackage{cite}
\usepackage{amsmath,amssymb,amsfonts}
\usepackage{graphicx}
\usepackage{textcomp}
\def\BibTeX{{\rm B\kern-.05em{\sc i\kern-.025em b}\kern-.08em
    T\kern-.1667em\lower.7ex\hbox{E}\kern-.125emX}}

\input{newcmd}

\usepackage[numbers,sort&compress]{natbib}
\usepackage{dblfloatfix}

\makeatletter
\g@addto@macro\ps@IEEEtitlepagestyle{%
  \def\@oddhead{\hfil\normalfont\normalsize 2026 IEEE 37th International Symposium on Software Reliability Engineering (ISSRE)\hfil}%
  \let\@evenhead\@oddhead}
\makeatother

\begin{document}
\bstctlcite{IEEEtranBSTctl}
\title{\ours: One Step Toward Test-Driven Development for
Repository-Level Code Generation}

\author{
  Yiran Hu,
  Shanchao Liang,
  Nan Jiang$^\dagger$,
  Yi Wu,
  and Lin Tan \\
  Purdue University \\
  {\small \{hu954, liang422, wu1827, lintan\}@purdue.edu},
  {\small $^\dagger$nan.nathan.jiang@gmail.com}
}

\maketitle

\begin{abstract}
\input{sections/abstract}
\end{abstract}


\input{sections/introduction}

\input{sections/related_work}

\input{sections/methodology}

\input{sections/experiment_setup}
\input{sections/evaluation}

\input{sections/conclusion}

\section*{Acknowledgment}
This research was supported in part by NSF 1901242 and 2006688. Any opinions, findings, and conclusions in this paper are those of the authors only and do not necessarily reflect the views of our sponsors.

\bibliographystyle{IEEEtranN}
{
\footnotesize
\bibliography{ref}
}

\end{document}

%% file: newcmd.tex
\newcommand{\todoc}[2]{{\textcolor{#1}{\textbf{#2}}}}

\definecolor{lightred}{RGB}{255, 204, 204}

\renewcommand{\todoc}[2]{\relax}

\newcommand{\acr}{AutoCodeRover\xspace}
\newcommand{\repocod}{{\textsc{RepoCod}}\xspace}
\newcommand{\code}[1]{\lstinline[basicstyle=\ttfamily\small,breaklines=true]@#1@}

\newcolumntype{a}{>{\columncolor{lightgray}}c}

\providecommand{\circled}[1]{\Circled[inner xsep=2.5pt,inner ysep=2.5pt]{#1}}

\DeclareRobustCommand{\ours}{%
    \ifmmode
        \text{T\scalebox{0.8}{ENET}}%
    \else
        T\scalebox{0.8}{ENET}%
    \fi
    \xspace
}

\newcommand{\distance}{10pt}

%% file: sections/abstract.tex
Test-Driven Development (TDD) is a widely adopted practice that requires developers to create and execute tests alongside implementation. With recent advances in Large Language Models (LLMs), developers can shift from manually writing the code to defining tests as executable specifications and delegating code synthesis to AI agents. However, enabling repository-level TDD under developer-written tests is challenging, requiring: (1) \textbf{specification enhancement}: identifying a concise yet representative test subset from large suites with rich task semantics; (2) \textbf{retrieval augmentation}: using tests to guide reasoning and context retrieval; and (3) \textbf{test-driven refinement}: interpreting test feedback for iterative improvement. We propose \ours, an agentic framework for repository-level code generation under the TDD paradigm. \ours includes: (1) a \emph{\textbf{test harness mechanism}} that selects a concise test suite considering both the caller diversity and invocation proximity; (2) a \emph{\textbf{tailored agent toolset}} for efficient retrieval and debugging; and (3) a \emph{\textbf{reflection-based refinement workflow}} that iteratively analyzes failures and updates implementations. 
\ours consistently outperforms the strongest baselines across backbones, achieving 69.08\% and 81.77\% Pass@1 on \repocod and RepoEval with Claude Sonnet 4, improving by 9.49 and 2.17 percentage points, respectively. Additionally, we present the first systematic study of how test suite characteristics influence LLM agent performance in TDD settings.

%% file: sections/introduction.tex
\begin{figure*}[t]
    \centering
    \includegraphics[width=0.9\linewidth]{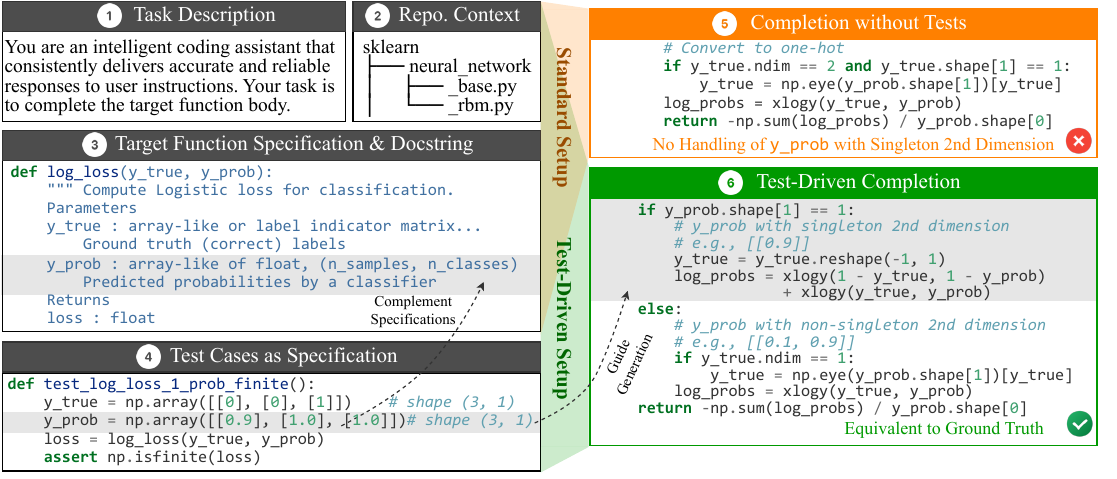}
    \caption{Examples of repository-level code generation under standard and test-driven setups.}
    \label{fig:motivation-example}
\end{figure*}

\section{Introduction}
\label{Intro}
Test-Driven Development is a widely adopted practice in software engineering that tightly couples the testing and implementation processes~\citep{TDD-def}: developers typically start by writing test cases that specify the desired behavior or capture potential failure scenarios, then incrementally implement and refine code to satisfy these tests. Extensive studies have demonstrated that TDD improves implementation accuracy, code quality, and design clarity~\citep{TDD4CodeGen, LLM4TDD, miu-fix, show_why_tdd_is_good1, show_why_tdd_is_good2, show_why_tdd_is_good3, show_why_tdd_is_good4, show_why_tdd_is_good5, GAI4-TDD, show_why_tdd_is_good6}.

In the era of AI-assisted  coding~\citep{ai-assisted-coding}, developers increasingly delegate code writing to large language models by providing high-level intentions. However, such descriptions often fail to capture detailed functionality, making it challenging for LLMs to implement correctly and sometimes even amplifying the ambiguity, especially with the repository-level context. This makes TDD even more crucial, as it systematically specifies and validates requirements through executable test cases to improve functional correctness. Recent industry practices further highlight the effectiveness of TDD in agentic coding workflows~\citep{ClaudeCodeBestPractices, ClaudeCodeTalk}.

Fig.~\ref{fig:motivation-example} shows how an LLM agent generates the target function for \code{scikit-learn_304} from \repocod~\citep{Benchmark_RepoCod}. In the standard setting, the agent is given \circled{1} the task description, \circled{2} repository context, and \circled{3} function specification. Based on these, it assumes that \code{y\_prob.shape[1]} (the size of distributions) is always greater than 1, since classification should involve at least two classes, and produces the implementation in \circled{5}. However, \code{scikit-learn} also requires supporting singleton probability representations for binary classification (e.g., \code{[0.9]} instead of \code{[0.1,0.9]}), which cannot be inferred from \circled{1}--\circled{3} and the implementation in \circled{5} fails on such inputs. In contrast, the TDD setting additionally provides \circled{4} test cases that include such usage examples (the gray lines in \circled{4}) for specification enhancement, guiding the agent to generate the correct implementation in \circled{6}.

While this highlights TDD as a promising setting, existing studies on automated code generation in TDD are largely limited to standalone function generation \citep{miu-fix, TDD4CodeGen, LLM4TDD, TiCoder, AceCoder}. Automating TDD in repository-level code generation remains underexplored. Although the ultimate goal of AI-driven TDD involves authoring new tests and the full "test-implement-refactor" cycle, a crucial step forward is enabling agents to reliably utilize and satisfy \textbf{existing developer-written} tests. This focused setting introduces several challenges.

First, unlike standalone functions, where tests directly invoke the target, software repositories contain large test suites where tests reach the target function through long and indirect call chains. How to strategically leverage these tests for \textbf{specification enhancement} to guide LLMs under computational budgets remains an open question.
Second, relevant context is often scattered across the repository, and implementations must follow repository-specific structures, imports, and conventions that are invisible at function granularity. This poses a key challenge for \textbf{retrieval augmentation}: effectively leveraging test context and signals to guide agents toward effective reasoning and retrieval.
Finally, test execution feedback is often sparse and noisy, and failures may arise from interactions far from the generated function, making it difficult for models to precisely interpret for \textbf{code refinement}.

To tackle these challenges, we introduce \ours, an agentic framework for repository-level code generation focusing on how to effectively utilize developer-written tests, retrieve informative context, and leverage test feedback for code refinement. The framework features three components. First, given developer-written test cases, \ours employs a \textbf{\emph{test harness mechanism (THM)}} that utilizes dynamic analysis to select a subset of tests that invoke target functions from distinct caller functions in the call stack to maximize coverage of diverse usage scenarios.
Second, it leverages a \textbf{\emph{tailored agent toolset}} for repository-level code generation, enabling efficient discovery of diverse usage patterns of the target function and supporting interactive debugging.
Third, it implements a \textbf{\emph{reflection-based refinement workflow (RRW)}} that iteratively improves generated code via failure analysis, contextual replenishment, and execution-based debugging. Together, these components allow \ours to generate more accurate code in complex real-world repositories. The core contributions are summarized as follows.
\begin{itemize}[left=0pt, itemsep=0pt, topsep=0pt, parsep=0pt, partopsep=0pt]
    \item We propose an effective test harness mechanism for efficient test-driven supervision, using dynamic analysis to select a concise subset of tests for specification enhancement.
    \item We design a tailored agent toolset that supports LLM agents to efficiently utilize test signals and perform context retrieval with interactive debugging.
    \item We build a reflection-based refinement workflow for test-aware debugging, enabling LLM agents to precisely interpret test feedback for code refinement. 
    \item We develop \ours, a test-driven agentic framework that integrates the above components for repository-level code generation. It achieves 69.08\% and 81.77\% Pass@1 on \repocod and RepoEval, outperforming the strongest agentic baselines by 9.49 and 2.17 percentage points (pp), respectively. These margins further widen to 20.10 and 6.97 pp under the DeepSeek-V3 backbone.
    \item We conduct the first systematic study of test-driven code generation using LLM agents at the repository level. The key findings include:
    \begin{itemize}[left=0pt, itemsep=0pt, topsep=0pt, parsep=0pt, partopsep=0pt]
        \item Too many tests do not necessarily lead to superior results; a moderate number, typically three to five representative tests, often yields optimal performance (Section~\ref{sec:RQ3}).
        \item Test suites considering both caller diversity and invocation proximity provide the most effective guidance for LLM agent repository-level code generation (Section~\ref{sec:RQ4}).
        \item While incorporating test signals in both context retrieval and refinement consistently improves the performance, it comes with higher token consumption. We should balance the trade-off between accuracy and efficiency (Section~\ref{sec:RQ5}).
    \end{itemize}
\end{itemize}

Our implementation and results are available at \url{https://github.com/lt-asset/tenet/}.

%% file: sections/related_work.tex
\begin{figure*}[t]
    \centering
    \includegraphics[width=\linewidth]{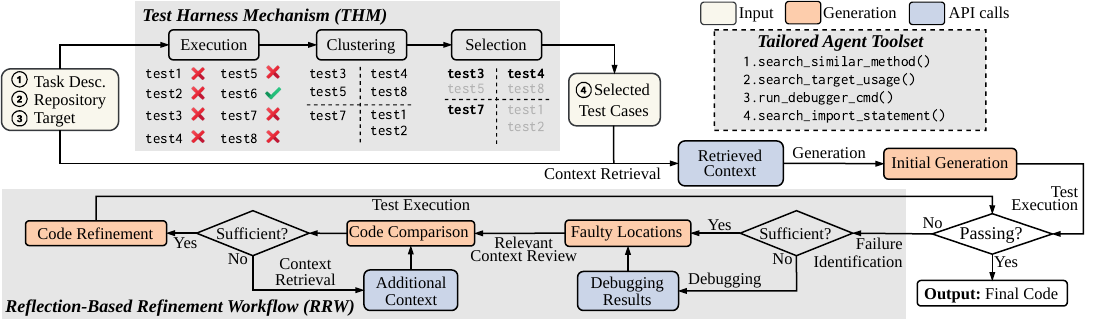}
    \caption{\ours workflow.}
    \label{fig:workflow}
\end{figure*}

\section{Related Work}
\subsection{Test-Driven Development}
A typical Test-Driven Development (TDD) workflow follows a “test–implement–refactor” cycle: developers begin with test cases that specify desired functionality, write the minimal code required to pass the tests, and refactor the implementation while ensuring tests continue to pass. Prior work has demonstrated the effectiveness of TDD in improving software quality~\citep{show_why_tdd_is_good1, show_why_tdd_is_good2, show_why_tdd_is_good3, show_why_tdd_is_good4, show_why_tdd_is_good5, show_why_tdd_is_good6} and has further explored its application to LLM-based code generation, primarily on standalone function generation benchmarks~\citep{TDD4CodeGen, LLM4TDD, miu-fix, TiCoder, AceCoder}. Under this setting, models do not need to navigate complex file dependencies or isolate the specific subset of tests relevant to a change.

In contrast, \ours takes a step towards repository-level TDD. Our setting requires the agent to select, utilize, and satisfy the \textbf{existing developer-written} test cases in the repository\citep{use-old-tests, liu2026issueexec}. It introduces new challenges in test selection for specification enhancement and test-guided retrieval that standalone baselines do not address.

\subsection{Repository-Level Code Generation}
Many code generation benchmarks focus on generating isolated code units~\citep{HumanEval, MBPP, APPS}, which differ from real-world software development where implementations must integrate with large repositories and satisfy project-specific dependencies. To better reflect such scenarios, repository-level code generation benchmarks have been proposed~\citep{Benchmark_RepoCod, repocoder_repoeval, CrossCodeEval, DevEval, codereval, CodeRAG-Bench, FEA-Bench, ExecRepoBench}, requiring models to implement new functionalities within complex codebases while maintaining consistency. 

Existing approaches for repository-level code generation with LLM can be categorized into non-agentic and agentic methods. Non-agentic approaches mainly follow two paradigms: Retrieval-Augmented Generation (RAG), which retrieves relevant repository context using semantic or structural strategies~\citep{repocoder_repoeval, RLCoder, REPOFORMER, RLPG, A3_CodGen, DraCo, GraphCoder, CodeRAG, REPOFUSE, RepoGraph, RepoHyper}; and feedback-driven refinement, which improves generated code using external signals such as compiler errors~\citep{CoCoGen} or symbolic planning~\citep{CodePlan}.
Agentic approaches formulate repository-level code generation as a multi-step decision-making process, where the model interacts with external tools to explore the repository, modify code, and validate changes~\citep{CodeAgent, OpenHands, SWE-agent, SpecRover}.

\vspace{-2pt}
\subsection{Large Language Model Agents}
LLM agents leverage LLMs as central reasoning engines to decompose complex problems, formulate multi-step plans, and interact with external environments through tool invocation. 
Prior work explores diverse agent designs for software engineering tasks.
Approaches including AutoCodeRover~\citep{ACR}, SpecRover~\citep{SpecRover}, as well as CodeAgent~\citep{CodeAgent}, are designed for specific scenarios, such as issue fixing or repository-level code generation. These methods typically rely on pre-defined workflows and toolsets to guide program analysis and code modification.
General-purpose software engineering agents, including SWE-Agent~\citep{SWE-agent}, CodeAct~\citep{CodeAct}, and Live-SWE-Agent~\citep{live-swe-agent}, can solve real-world GitHub issues by equipping LLMs with tools for code navigation, editing, and test execution, with Live-SWE-Agent further enabling agents to dynamically construct their own toolsets on the fly. 

However, when generating code in a test-driven manner, they lack mechanisms to select informative tests and to effectively leverage execution feedback for refinement. \ours{} addresses this by using THM to select diverse and informative test cases for better specification, equipping LLMs with a powerful toolset for retrieval and interactive debugging, and adopting RRW for structured and effective code refinement.

%% file: sections/methodology.tex
\section{Approach: \ours}
\label{sec:approach}
Fig.~\ref{fig:workflow} illustrates the workflow of \ours{}, which consists of three major components: (1) the \emph{test harness mechanism (THM)}, (2) the \emph{tailored agent toolset}, and (3) the \emph{reflection-based refinement workflow (RRW)}. The THM is a dynamic-analysis preprocessing stage that operates \emph{outside} the agent: it first selects a small yet effective test subset from the repository's complete test suite. This subset is provided as additional input \circled{4}, alongside the \circled{1} task description, \circled{2} repository context, and \circled{3} target function specification. Throughout the subsequent workflow, the agent can access and execute only the selected tests, while all remaining tests are held out from the agent. Guided by the test subset, \ours{} explores the repository using the tailored agent toolset, which supports efficient retrieval and interactive debugging. After gathering sufficient context, it generates a solution and validates it against the selected tests. If any test fails, \ours{} enters the RRW to repair the generated code.

During the RRW, the agent assesses whether the current information provides adequate support for a correct refinement, including fault localization and issue understanding. If not, the RRW prompts the agent to retrieve additional context and use interactive debugging to gather further insights. This loop continues until the agent deems the evidence sufficient for a fix, at which point it generates a candidate refinement. The updated code is then validated using the THM-selected tests. \ours stays in the RRW until all selected tests pass or a maximum number of attempts is reached.

\subsection{Test Harness Mechanism}
\label{sec:test harness}
Test cases can expose diverse usage scenarios of the target function, serving as a form of specification enhancement beyond the function signature and docstring. However, dumping the entire test suite to the LLM is infeasible due to two reasons.
\begin{itemize}[left=0pt, itemsep=0pt, topsep=0pt, parsep=0pt, partopsep=0pt]
     \item \textbf{Effectiveness}. Providing massive test cases can overwhelm the LLM with long, complex prompts, increasing the cognitive load of the model~\citep{TDD4CodeGen,lost_in_middle}.
    \item \textbf{Efficiency}. Such massive test suites and long prompts inevitably incur higher generation latency and execution workload, reducing the efficiency of the agent pipeline~\citep{Inheritance_test}.
\end{itemize}

To identify a concise and effective test subset from the entire repository test suite, \ours{} employs an effective \emph{THM} that first executes the full test suite against the unimplemented target function to collect failing cases. These failing cases are then clustered by the \emph{caller} function that directly invokes the target function in their call stack. The intuition is that different caller functions are likely to expose distinct usage patterns and test diverse aspects of the target function's logic, thereby providing complementary test coverage. For the example shown in Fig.~\ref{fig:motivation-example}, the test function with call chain \underline{\code{test\_log\_loss\_1\_prob\_finite}} $\rightarrow$ \code{log\_loss} tests the target function under the binary classification scenario and covers the \code{if} branch, while another cluster of test functions, with call chain $\ldots\rightarrow$ \underline{\code{\_backprop}} $\rightarrow$ \code{log\_loss}, all test the target function by covering the \code{else} branch.

From these clusters, \ours{} selects at most $T$ test cases. The selection further prioritizes tests with the shortest call chain from the entry test function to the target function. 
The rationale is that tests with shorter invocation paths may expose cleaner execution contexts and more localized failure signals. In contrast, tests with deeper invocation paths often provide longer and more complex feedback, indicating complicated execution paths and error propagation. Selecting tests with shorter call chains may expose the agent to more concise and focused feedback, thereby simplifying the refinement process and reducing the overall task difficulty.

In practice, the process first attempts to pick one representative test case from each cluster with the shortest call chain. If the number of clusters is greater than or equal to $T$, the top $T$ clusters are chosen, with one test case selected from each. If the number of clusters is fewer than $T$, additional test cases with the shortest call chains are selected until the budget of $T$ is reached. This strategy ensures that the final subset of test cases remains both diverse across failure patterns and closely tied to the target function's behavior.

We emphasize that the full-suite execution above is performed by the THM alone for candidate discovery; the subsequent agent only has access to the selected tests.
We set $T=3$ for the algorithm described above based on preliminary experiments on the sphinx project from \repocod (Section~\ref{sec:RQ3}). We study alternative test selection strategies in Section~\ref{sec:RQ4}.

\subsection{Tailored Agent Toolset}
\label{sec:toolset}
\ours{} provides a tailored toolset that extends the abstract-syntax-tree (AST)-based toolset of SpecRover~\citep{SpecRover} for structural context retrieval and interactive debugging. The API toolsets of existing agents can be roughly divided into two categories: AST-based, such as \acr{} and SpecRover, and terminal-command based, such as SWE-Agent~\citep{SWE-agent} and OpenHands~\citep{OpenHands}. While AST-based interfaces allow structural navigation and terminal-command based interfaces offer more flexibility, they still face several limitations when generating functions under repository-level context.

First, RAG techniques can substantially improve generation accuracy by providing relevant code examples. However, existing LLM agents have not incorporated semantic retrieval as an API beyond basic repository navigation, which usually requires multiple attempts to locate the desired context. Second, understanding the use cases is crucial for accurate code generation. Yet, existing LLM agents rely on terminal commands to find the usage of a certain function based on string matching, which is inefficient and error-prone. Third, although interactive debugging plays an important role in refining code~\citep{debug_gym}, existing agents treat it as an end-to-end process, lacking support for fine-grained interactive debugging that enables stepwise evidence collection and fault diagnosis. 

To address these, we extend SpecRover's AST-based toolset with four new APIs:
\begin{itemize}[left=0pt, itemsep=0pt, topsep=0pt, parsep=0pt, partopsep=0pt]
    \item \textbf{\code{search\_similar\_method(n)}} retrieves the top-\code{n} methods most relevant to the signature and docstring of the target function, ranked by BM25 similarity~\citep{Benchmark_RepoCod, CodeAgent}. This API enables the agent to collect context as references efficiently.
    
    \item \textbf{\code{search\_target\_usage(n)}} 
    retrieves up to n usage examples of the target function via AST analysis. Unlike keyword-matching commands that require multiple indirect queries and often return noisy snippets, this API provides usage contexts in a single step, making it easier for the agent to understand how the target function is invoked.
    
    \item \textbf{\code{run\_debugger\_cmd(cmd)}} executes terminal commands within a container session, including interactive debugger commands (e.g., \code{pdb}) for step-by-step execution, variable inspection, and stack frame traversal.

    \item \textbf{\code{search\_import\_statement(f)}} retrieves all top-level import statements in the specified file \code{f}. This enables the agent to analyze dependencies and disambiguate call paths in cross-file analysis.
\end{itemize}
Together, these APIs address the identified limitations by allowing \ours{} to efficiently retrieve relevant context and conduct structured interactive debugging.

\subsection{Reflection-Based Refinement Workflow}
\label{sec:refinement}
With the THM and the tailored toolset, \ours{} first generates a code snippet which is then validated against the chosen test cases. If any THM-selected test fails, \ours{} enters the RRW to revise the code snippet iteratively. Throughout the RRW, generation and refinement are validated \emph{exclusively} against the THM-selected tests.
Unlike refinement in the setting of self-contained functions, where the LLM only needs to review and revise its own faulty implementation~\citep{google-self-repair, mit-self-repair, agentcoder-self-repair}, the generation of incorrect code within a repository could be due to more complex reasons. For example, LLM can get “lost in the middle” due to long trajectories~\citep{lost_in_middle}, overlooking important context such as the usage of repository-specific functions.

To address such issues, \ours{} employs a structured and bounded \emph{RRW} rather than relying on unguided self-reflection. Each refinement attempt follows three stages:
(1) \textbf{Failure Identification.} The LLM identifies potentially faulty locations from the current trajectory and invokes debugger commands only when additional execution evidence is needed.
(2) \textbf{Context Review.} The LLM is then prompted to identify relevant code snippets (e.g., implementations of similar functionality) based on the current trajectory to avoid redundant tool calling. When such snippets are available, the LLM is guided to explain their implementations, such as the handling of edge cases or the usage of specific functions, and compare them with the faulty implementations to extract insights for bug fixing.
(3) \textbf{Code Refinement.} The LLM determines whether the available evidence is sufficient for refinement. If so, it formulates and applies a concrete repair strategy; otherwise, it selectively gathers additional context or debugging signals before revising the solution.
The RRW terminates when either all THM-selected tests pass or the refinement budget is exhausted, preventing indefinite oscillation among failing solutions.

\input{tables/RQ1-deepseek}

%% file: tables/RQ1-deepseek.tex
\begin{table*}[!t]
    \centering
    \scriptsize
    \setlength{\tabcolsep}{2.8pt}
    \renewcommand{\arraystretch}{0.8}
    
    \caption{Comparison of Pass@1, token consumption, monetary cost, and Cost-of-Pass (CoP) on \repocod and RepoEval under Claude Sonnet 4 and DeepSeek-V3.}
    \begin{adjustbox}{max width=1\textwidth}
    \begin{tabular}{ll|rrrrr|rrrrr}
    \toprule
    \multirow{2}{*}{Backbone} & \multirow{2}{*}{Approach}
    & \multicolumn{5}{c|}{\repocod}
    & \multicolumn{5}{c}{RepoEval} \\
    \cmidrule(lr){3-7} \cmidrule(lr){8-12}
    & 
    & Pass@1 (\%)$\uparrow$
    & Input $\downarrow$
    & Output $\downarrow$
    & Cost (\$)$\downarrow$
    & CoP $\downarrow$
    & Pass@1 (\%)$\uparrow$
    & Input $\downarrow$
    & Output $\downarrow$
    & Cost (\$)$\downarrow$
    & CoP $\downarrow$ \\
    \midrule

    \multirow{5}{*}{Sonnet 4}
    & RepoCoder
    & 26.22 & 14,225 & 787 & \$0.03 & 0.11
    & 53.08 & 6,048 & 306 & \$0.01 & 0.02 \\
    & SpecRover
    & 33.33 & 95,884 & 6,609 & \$0.35 & 1.05
    & 72.12 & 52,317 & 4,643 & \$0.20 & 0.28 \\
    & OpenHands
    & 59.18 & 1,119,149 & 8,988 & \$0.64 & 1.08
    & 79.60 & 471,103 & 4,434 & \$0.30 & 0.38 \\
    & SWE-Agent
    & 59.59 & 597,507 & 1,252 & \$0.45 & 0.76
    & 67.02 & 344,374 & 1,242 & \$0.30 & 0.45 \\
    & \ours
    & \textbf{69.08} & 194,932 & 6,560 & \$0.53 & 0.76
    & \textbf{81.77} & 111,934 & 3,790 & \$0.28 & 0.34 \\
    \midrule

    \multirow{5}{*}{DeepSeek-V3}
    & RepoCoder
    & 19.80 & 12,477 & 511 & \$0.004 & 0.018
    & 48.26 & 12,830 & 222 & \$0.003 & 0.007 \\
    & SpecRover
    & 18.24 & 61,378 & 5,456 & \$0.023 & 0.124
    & 46.11 & 46,603 & 4,256 & \$0.017 & 0.038 \\
    & OpenHands
    & 29.08 & 353,648 & 6,608 & \$0.051 & 0.176
    & 43.16 & 176,073 & 2,210 & \$0.025 & 0.059 \\
    & SWE-Agent
    & 27.04 & 386,932 & 736 & \$0.049 & 0.181
    & 35.66 & 70,540 & 533 & \$0.007 & 0.020 \\
    & \ours
    & \textbf{49.18} & 147,968 & 4,645 & \$0.026 & 0.052
    & \textbf{55.23} & 82,473 & 2,967 & \$0.017 & 0.031 \\
    \bottomrule
    \end{tabular}
    \end{adjustbox}
    \label{table:baseline-compare}
\end{table*}

%% file: sections/experiment_setup.tex
\section{Experimental Setup}
\label{sec:exp_setup}

\subsection{Research Questions}
To assess the effectiveness of \ours and to gain deeper insights into repository-level code generation under the TDD setting, we formulate the following research questions (RQs).
\begin{itemize}[left=0pt, itemsep=0pt, topsep=0pt, parsep=0pt, partopsep=0pt]
    \item \textbf{RQ1:} How effective is \ours compared to other repository-level code generation baselines?
    \item \textbf{RQ2:} How does each component of \ours contribute to its overall effectiveness?
    \item \textbf{RQ3:} What effect does the quantity of test cases in \ours have on code generation performance? 
    \item \textbf{RQ4:} How do different test selection strategies affect the code generation pass rate?
    \item \textbf{RQ5:} What is the impact of using test cases at different stages of \ours on performance?
\end{itemize}

\subsection{Benchmarks}
We evaluate our approach on two repository-level code generation benchmarks, \repocod~\citep{Benchmark_RepoCod} and the function-level tasks of RepoEval~\citep{repocoder_repoeval}. Both benchmarks preserve full repository context and executable test suites, comprising 980 and 373 function generation tasks, respectively. Each task requires synthesizing a missing target function such that the completed implementation passes the associated tests. Although both benchmarks fall under repository-level code generation, they differ in both the scope of required code modifications and the size of test suites. \repocod tasks typically involve synthesizing more substantial code, with an average of 38.18 edited lines of code (LOC) and 334.42 code-edit tokens\footnote{The average number of code-edit tokens per task is computed using a fixed byte-pair encoding tokenizer from the \texttt{tiktoken} library.}. Each task is evaluated against approximately 68 tests\footnote{This count does not distinguish test functions executed with different inputs. When each input variant is counted as a separate test, the average increases to 313~\citep{Benchmark_RepoCod}.} to determine Pass@1. In contrast, RepoEval tasks generally require more concise implementations, averaging 9.78 edited LOC and 84.92 code-edit tokens, with approximately 15 evaluation tests per task. This distinction enables evaluating \ours across varying implementation scopes. For each task, we execute all agentic approaches in isolated Docker environments with complete dependencies to ensure consistent and reproducible evaluation.

A related but distinct line of work focuses on automated issue fixing, exemplified by SWE-Bench Verified~\citep{swe-bench-M, swesmith, swe-bench}. In this setting, the primary objective is software maintenance: agents interpret issue descriptions, localize faulty code, and generate minimal patches to repair existing implementations. Prior analysis shows that 87.20\% of SWE-Bench Verified tasks correspond to bug fixing, while only 8.60\% involve feature implementation~\citep{swe-poly-bench}. This distribution reflects a fundamental difference in task scope. Unlike issue-fixing benchmarks that emphasize fault localization and patch generation, the setting we consider requires synthesizing new functionality under test-driven constraints, introducing distinct challenges in test selection, context understanding, and test-guided refinement.

\subsection{Baselines}
Across all RQs, we use DeepSeek-V3~\citep{deepseekv3-technical-report} as the primary backbone for its cost efficiency. In RQ1, we compare \ours{} with four open-source baselines covering both non-agentic and agentic paradigms, and further evaluate all methods with Claude Sonnet 4~\citep{Claude} as a high-performance reference.
All baselines receive the task description and full repository context, including the complete test suites. Agentic baselines may freely inspect and execute any repository test. In contrast, \ours{} exposes only the THM-selected tests to the downstream agent; all other tests are withheld during generation and refinement and used only for final evaluation. We configure \ours{} with up to three THM-selected tests, 15 pre-generation conversation rounds, five RRW refinement attempts, and 15 conversation rounds per refinement attempt for debugging and context retrieval. For all baselines and configurations, we set the model temperature to 0 and sample one patch per task.
\begin{enumerate}[wide, itemsep=0pt, topsep=0pt, parsep=0pt, partopsep=0pt]
    \item \textbf{RepoCoder} adopts an iterative retrieval–generation framework~\citep{repocoder_repoeval}. We follow the original configuration on RepoEval. For \repocod{}, we use a 12,288-token retrieval window, retrieve up to 30 snippets per query, and cap the maximum completions at 4,096 tokens.
    \item \textbf{SpecRover}~\citep{SpecRover} is a multi-agent framework coordinating specialized agents for retrieval, generation, testing, and reflection. We follow the same configuration in the paper.
    \item \textbf{SWE-Agent} equips LLMs with shell commands and custom actions~\citep{SWE-agent} for general software engineering tasks. We use a maximum of 50 conversation rounds for each task.
    \item \textbf{OpenHands} is an open-source platform for developing software engineering agents~\citep{OpenHands}. We employ the default CodeAct Agent~\citep{CodeAct}, which supports shell command execution, file reading, and file editing. OpenHands follows the same conversation rounds as SWE-Agent.
\end{enumerate}

The round budgets for SWE-Agent and OpenHands are relatively higher, as they support only one tool invocation per response, while \ours can perform multiple API calls in a single round. This adjustment balances comparable tool-calling capacity and cost efficiency.

%% file: sections/evaluation.tex
\section{Evaluation Results}
\label{sec:eval}

\input{sections/evaluation/RQ1}

\input{sections/evaluation/RQ2}

\input{sections/evaluation/RQ3}

\input{sections/evaluation/RQ4}

\input{sections/evaluation/RQ6}

%% file: sections/evaluation/RQ1.tex
\subsection{RQ1: Baseline Comparison}
\label{sec:RQ1}

We report the performance of \ours{} and baseline approaches on \repocod and RepoEval under two backbones in Table~\ref{table:baseline-compare}, including Pass@1~\citep{HumanEval}, average token consumption, average monetary cost, and Cost-of-Pass (CoP)~\citep{Cost-of-Pass}. CoP is defined as the expected monetary cost to obtain one correct solution for a problem, computed as the ratio between the average monetary cost per task and the Pass@1 of an approach. Lower CoP indicates a more favorable accuracy-cost trade-off.

\textbf{Accuracy Gains.} \ours achieves the highest Pass@1 across \emph{both} benchmarks under \emph{both} backbones. Under Sonnet 4, it reaches 69.08\% and 81.77\% Pass@1, surpassing the strongest baselines by 9.49 and 2.17 pp, respectively. Under DeepSeek-V3, \ours attains 49.18\% and 55.23\%, where its margin over the strongest baselines \emph{widens} to 20.10 and 6.97 pp. This demonstrates the effectiveness of our design.

\textbf{Cost Efficiency.} \ours achieves a favorable accuracy--cost tradeoff under both backbones.
\begin{enumerate}[wide, itemsep=0pt, topsep=0pt, parsep=0pt, partopsep=0pt]
    \item \textit{Token Consumption.} Agentic baselines incur substantially higher token usage compared with RepoCoder. On \repocod under Sonnet 4, OpenHands and SWE-Agent consume 1,119,149 and 597,507 input tokens per task on average, corresponding to 574.12\% and 306.52\% of \ours, respectively. A similar trend is observed on RepoEval (420.88\% and 250.46\% of \ours's input tokens) and under DeepSeek-V3 (239\% and 261\% on \repocod). This overhead primarily stems from two factors: (1) lengthy system prompts that encode detailed execution policies, security constraints, and multi-step workflows; and (2) reliance on terminal-level commands (e.g., \code{grep}, \code{find}) that require many sequential, fragmented steps for code retrieval. These interactions cause trajectories to grow rapidly, resulting in a dramatic increase in token consumption.
    \item \textit{Monetary Cost.} Under Sonnet 4 on \repocod, \ours attains a higher Pass@1 while maintaining a comparable CoP (0.76) to SWE-Agent, which achieves the strongest Pass@1 and CoP among agentic baselines; on RepoEval, it achieves the highest Pass@1 with a competitive CoP of 0.34, only slightly higher than SpecRover, while outperforming it by 9.65 pp in Pass@1. The advantage is even clearer under the consistent DeepSeek-V3 backbone: \ours attains the \emph{lowest CoP among all agentic approaches} on \repocod, and a competitive CoP on RepoEval (0.031), second only to SWE-Agent, which trails \ours by 19.57 pp in Pass@1.
\end{enumerate}

\textbf{Overall}, \ours achieves high accuracy without sacrificing cost efficiency, and its advantage holds under different backbone models. While TENET excels, general-purpose agents exhibit competitive performance across benchmarks~\citep{GAIA, swe-poly-bench, openagentsafety, swe-bench-M}, reflecting their strong generality.

\begin{figure*}[!t]
    \centering
    \includegraphics[width=0.9\linewidth]{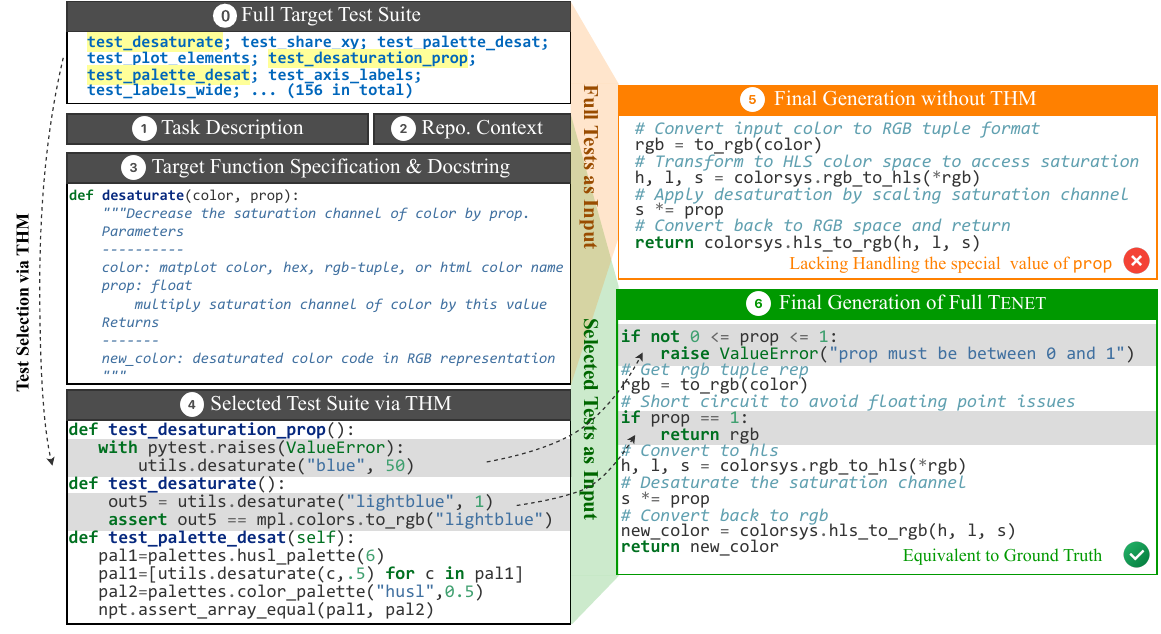}
    \caption{Case study on task \code{seaborn\_34} from \repocod, explaining how THM guides \ours toward correct code generation.}
    \label{fig:THM-example}
\end{figure*}

%% file: sections/evaluation/RQ2.tex
\subsection{RQ2: Contributions of \ours’s Components}
\label{sec:RQ2}
\input{tables/5_ablation}

To address RQ2, we first remove each component from the full \ours system to perform an ablation study on \repocod{}. Removing the THM (\ours{}$_{\text{-THM}}$) provides the agent with full target test suite instead of the selected subset; removing the tailored toolset (\ours{}$_{\text{-APIs}}$) limits the agent to the SpecRover toolset; and removing the RRW (\ours{}$_{\text{-RRW}}$) applies naive refinement given test feedback if test execution fails. Table~\ref{table:ablation} reports the Pass@1, average token consumptions (Columns ``Input $\downarrow$" and ``Output $\downarrow$"), and the number of API calls on \repocod. 
\emph{Across all cases, removing any component results in a clear decline in Pass@1.}

\subsubsection{Test Harness Mechanism}
Removing THM causes the largest Pass@1 drop (16.12 pp). The token consumption also rises substantially with inputs by 40.69\% and outputs by 19.41\%, along with a 45.53\% rise in API calls. This is mainly due to two factors. First, each \repocod{} task contains 68 test cases on average, far more than the curated suites from the THM. Feeding the model with the full suite introduces redundancy and noise, reducing accuracy. Second, handling the full suite requires analyzing more tests, which increases API calls and token usage.

Fig.~\ref{fig:THM-example} illustrates the task \code{seaborn\_34} from \repocod to demonstrate the effectiveness of the THM. The full target test suite of \code{seaborn\_34} contains 156 test cases for the target function \code{desaturate}, located in \code{seaborn/utils.py}. Without the THM, \circled{0} the large test suite overwhelms the LLM, causing the agent to overlook critical test signals and generate \circled{5} the incorrect code. Specifically, the generated function fails to satisfy the input validation requirements specified in \code{test\_desaturation\_prop}. It further fails the test \code{test\_desaturate} with the following error message.
\lstset{
  basicstyle=\ttfamily\scriptsize,     
  keywordstyle=\color{blue},      
  commentstyle=\color{gray},      
  stringstyle=\color{red!60!brown}, 
  breaklines=true,                
  frame=single,                   
  numbers=left,                  
  numberstyle=\tiny\color{gray},  
  showstringspaces=false,         
  captionpos=b,                  
}
\begin{lstlisting}[language={}, basicstyle=\ttfamily\scriptsize\color{black}, frame=single, numbers=none, breaklines=true]
E   AssertionError: assert (0.6784313725...9607843137256) == (0.6784313725...9607843137255) 
E     At index 1 diff: 0.8470588235294119 != 0.8470588235294118
\end{lstlisting}
The failure occurs because the incorrect code in \circled{5} always performs an RGB$\to$HLS$\to$RGB round-trip, even for the boundary case \code{prop==1}, introducing small floating-point deviations that cause the test to fail. In contrast, THM selects three representative test cases without overwhelming the agent, and \ours uses two refinement rounds to generate the correct code in \circled{6} that properly handles all edge cases.

\input{tables/RQ2-THM-ablation}

We further examine whether THM-selected tests benefit other agentic baselines beyond \ours{}. Keeping the RQ1 baseline setting unchanged, we additionally provide SWE-Agent with the THM-selected tests for each task and instruct it to verify its implementation against them. Table~\ref{table:sweagent-thm} reports Pass@1, average token consumption, and API calls on \repocod{}.
THM-selected tests improve SWE-Agent's Pass@1 by 13.98 pp, from 27.04\% to 41.02\%, at the cost of moderately higher token consumption and 25.19\% more API calls. This gain is largely driven by SWE-Agent's original test-usage behavior: despite unrestricted access to the full test suite, it submits patches without any test verification in 28.98\% of tasks. THM-selected tests, together with an explicit verification instruction, make testing a routine step in the workflow. \textbf{These results demonstrate that THM is an effective and transferable test-selection mechanism across agent designs.} However, SWE-Agent\textsubscript{+THM} still trails \ours{} by 8.16 pp, indicating that selected tests alone are insufficient; the tailored toolset and RRW are needed to fully exploit them.

\begin{figure}[t]
    \centering
    \includegraphics[width=\columnwidth,  trim = {1mm 1mm 0mm 1mm}, clip]{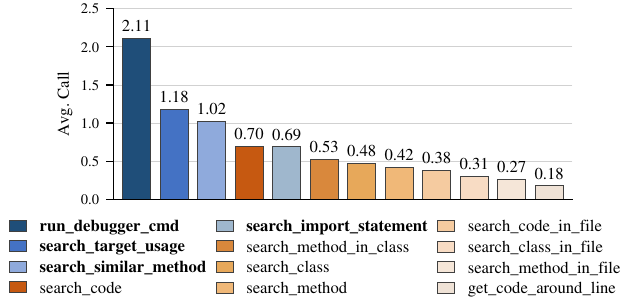}
    \caption{Average API calls per task on \repocod (\ours, DeepSeek-V3).}
    \label{fig:avg_api}
\end{figure}

\begin{figure*}[!t]
    \centering
    \includegraphics[width=\linewidth]{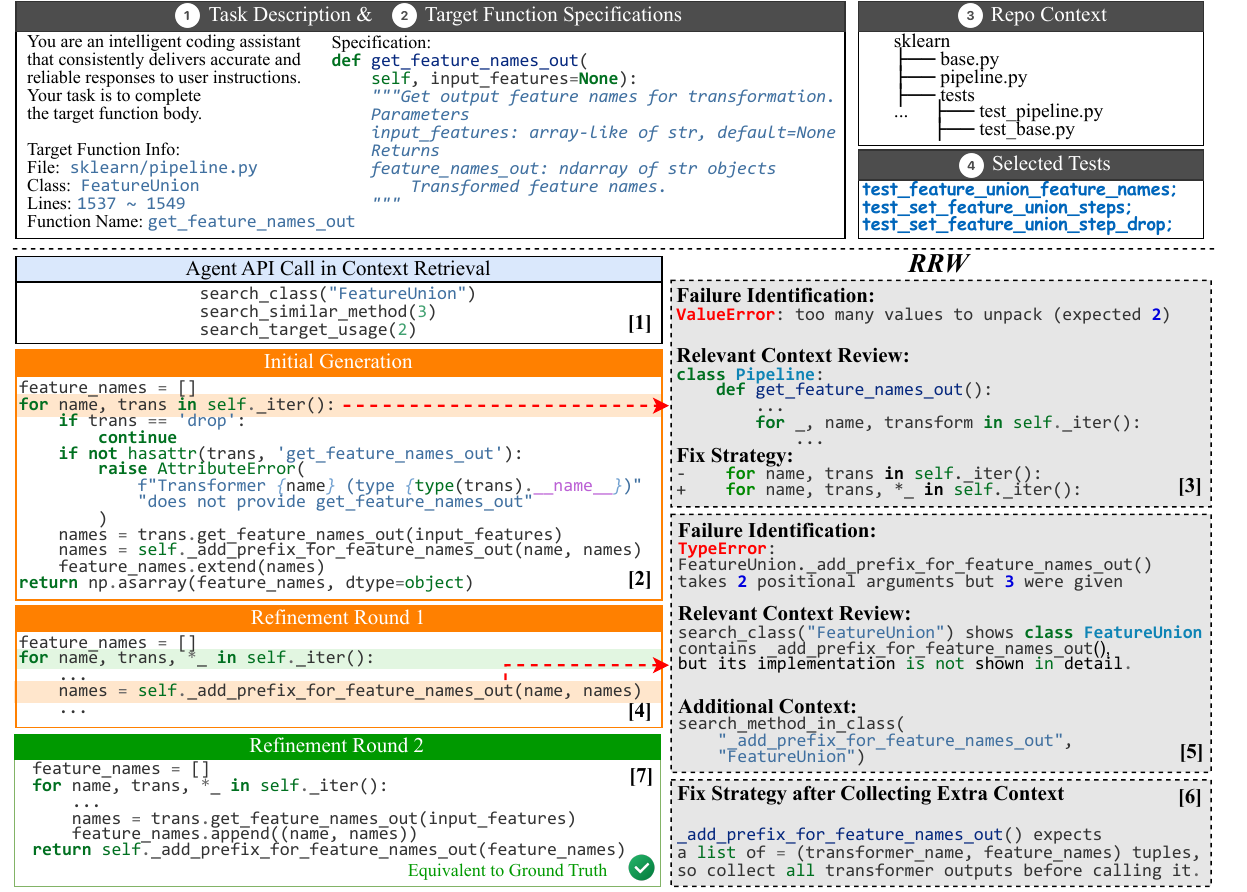}
    \caption{Case study on task \code{scikit\_49} in \repocod, explaining how the RRW contributes to the code refinement in \ours.}
    \label{fig:RRW-example}
\end{figure*}

\subsubsection{Tailored Agent Toolset}
Removing the tailored toolset reduces Pass@1 by 14.89 pp. Without \ours{'s} specialized tools, such as retrieving semantically similar code, the agent must rely on less efficient APIs to analyze the context, which increases reasoning complexity. Though the input tokens are fewer, the more expensive output tokens and API calls increase substantially. Fig.~\ref{fig:avg_api} shows the average call frequency of different APIs\footnote{The APIs not highlighted in bold are inherited from SpecRover.}. Our four newly introduced APIs (colored in blue) rank among the top five. We observe that \code{run\_debugger\_cmd} dominates with the highest frequency (2.11), highlighting the frequent usage of interactive debugging in the RRW. The next most frequently used APIs are \code{search\_target\_usage} (1.18) and \code{search\_similar\_method} (1.02), suggesting that retrieving usage examples and similar methods are also favorable in the TDD setting.
\textbf{These observations indicate that the model consistently favors our tailored toolset in the TDD setting, achieving efficient retrieval and improved performance with only a minor increase in input costs while reducing output costs and API overhead.}

\begin{figure}[t]
    \centering
    \includegraphics[width=0.8\columnwidth, trim = {0mm 3mm 0mm 2.2mm}, clip]{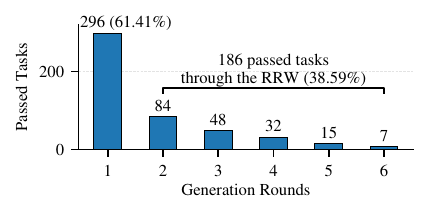}
    \caption{Number of solved tasks per round (\ours{}, DeepSeek-V3).}
    \label{fig:passround}
\end{figure}

\subsubsection{Reflection-Based Refinement Workflow}
Removing the RRW leads to a decrease of 9.24 pp on Pass@1. We let \ours{}\textsubscript{-RRW} naively regenerate code without explicit reasoning about fault localization or code comparison. As a result, the performance drops while the token consumption and API calls slightly decrease. Shown in Fig.~\ref{fig:passround}, solved tasks drop gradually over rounds, and out of 482 passed tasks in total, 296 (61.41\%) are solved in the first attempt. The remaining 186 tasks (38.59\%) are recovered through the RRW. 

Fig.~\ref{fig:RRW-example} illustrates the task \code{scikit_49} from \repocod to show how the RRW enables \ours{} to progressively correct generation errors and converge to the ground-truth solution. The task requires generating the \code{get_feature_names_out} method of the \code{FeatureUnion} class in \code{sklearn/pipeline.py}. \ours{} first invokes three APIs in \code{[1]} to retrieve the \code{FeatureUnion} class context, similar methods, and usage examples of the target function. Believing the collected context to be sufficient, \ours{} generates an initial implementation in \code{[2]}. However, test execution exposes a runtime error in \code{[3]}: \emph{ValueError: too many values to unpack}. Guided by the RRW, \ours{} identifies the failure location, reviews the retrieved context, locates a similar loop structure in the reference implementation, and produces the first refinement in \code{[4]}.

\input{tables/7_testnum_pass1}

\input{tables/12_strategies_full_repository_pass1}

Upon re-testing, a new error is triggered: \emph{TypeError: \_add\_prefix\_for\_feature\_names\_out takes 2 positional arguments but 3 were given}. \ours{} quickly localizes the failure and recognizes that the current trajectory only contains the signature of 
\code{\_add\_prefix\_for\_feature\_names\_out}, but not its full body: the earlier \code{search\_class("FeatureUnion")} call was truncated due to context-window constraints. This incomplete context causes the agent to invoke the method with an incorrect argument structure. Therefore, \ours{} issues an additional query, \code{search\_method\_in\_class("\_add\_prefix\_for\_feature\_names\_out", "FeatureUnion")}, to retrieve the complete method definition, as shown in \code{[5]}. With the correct usage clarified, \ours{} applies the fix in \code{[6]} and produces a final refinement in \code{[7]} that matches the ground-truth implementation.

\textbf{In summary, despite increased token usage, the RRW consistently enhances the effectiveness of \ours, by iteratively identifying failure signals, reviewing context, and adaptively invoking APIs when necessary.}

%% file: tables/5_ablation.tex
\begin{table}[t]
    \centering
    \scriptsize
    \setlength{\tabcolsep}{1pt}
    \caption{Ablation study of \ours{} using DeepSeek-V3 on \repocod{}. Compared with \ours, \textcolor{red}{red arrows} indicate worse results and \textcolor{blue}{blue arrows} indicate better results.}
    \vspace{-3pt}
    \begin{tabular}{l | lllr@{}l}
        \toprule
\textbf{Variants} & \textbf{Pass@1 (\%) $\uparrow$} & \textbf{Input $\downarrow$} & \textbf{Output $\downarrow$} & \multicolumn{2}{c}{\textbf{API Call $\downarrow$}}\\
    \midrule
    \ours{}\textsubscript{-THM} & 33.06\textsubscript{\textcolor{red}{$\downarrow 16.12$ pp}} & 208,179\textsubscript{\textcolor{red}{$\uparrow 40.69\ \%$}} & 5,547\textsubscript{\textcolor{red}{$\uparrow 19.41\ \%$}}  & 12.05&\textsubscript{\textcolor{red}{$\uparrow 45.53\ \%$}}\\
    \ours{}\textsubscript{-APIs} & 34.29\textsubscript{\textcolor{red}{$\downarrow14.89$ pp}} & 138,031\textsubscript{\textcolor{blue}{$\downarrow  6.72\ \%$}} & 6,358\textsubscript{\textcolor{red}{$\uparrow 36.87\ \%$}} &  10.53&\textsubscript{\textcolor{red}{$\uparrow 27.17\ \%$}} \\
    \ours{}\textsubscript{-RRW} & 39.94\textsubscript{\textcolor{red}{$\downarrow 9.24$ pp}} & 132,427\textsubscript{\textcolor{blue}{$\downarrow 10.50\ \%$}} & 4,008\textsubscript{\textcolor{blue}{$\downarrow 13.71\ \%$}} & 6.62&\textsubscript{\textcolor{blue}{$\downarrow 20.05\ \%$}}\\
    \midrule
    \ours{} & 49.18 & 147,968 & 4,645 & 8.28\\
    \bottomrule
    \end{tabular}
    \label{table:ablation}
\end{table}

%% file: tables/RQ2-THM-ablation.tex
\begin{table}[t]
    \centering
    \scriptsize
    \setlength{\tabcolsep}{1pt}
    \caption{Effect of THM-selected tests on SWE-Agent using DeepSeek-V3 on \repocod{}.}
    \vspace{-3pt}
    \begin{tabular}{l | lllr@{}l}
        \toprule
        \textbf{Variants}
        & \textbf{Pass@1 (\%) $\uparrow$}
        & \textbf{Input $\downarrow$}
        & \textbf{Output $\downarrow$}
        & \multicolumn{2}{c}{\textbf{API Calls $\downarrow$}} \\
        \midrule

        SWE-Agent
        & 27.04
        & 386,932
        & 736
        & 21.12 & \\

        SWE-Agent\textsubscript{+THM} 
        & 41.02\textsubscript{\textcolor{blue}{$\uparrow 13.98 $ pp}}
        & 412,377\textsubscript{\textcolor{red}{$\uparrow 6.58\ \%$}}
        & 916\textsubscript{\textcolor{red}{$\uparrow 24.46\ \%$}}
        & 26.44
        & \textsubscript{\textcolor{red}{$\uparrow 25.19\ \%$}} \\

        \bottomrule
    \end{tabular}
    \label{table:sweagent-thm}
\end{table}

%% file: tables/7_testnum_pass1.tex
\begin{table*}[!t]
    \centering
    \caption{Pass@1 (\%) of \ours with different test suite sizes on \repocod.}
    \setlength{\tabcolsep}{9pt}
    \renewcommand{\arraystretch}{0.9}
    \scriptsize
    \begin{tabular}{l | S | SSSSSSSSSS | S}
        \toprule
\textbf{\#Test} & \textbf{Sph.} & \textbf{Sea.} & \textbf{Fla.} & \textbf{Xar.} & \textbf{Sym.} & \textbf{More.} & \textbf{Data.} & \textbf{Skl.} & \textbf{Ast.} & \textbf{Pyl.} & \textbf{Plot.} & \textbf{Total} \\
    \midrule
    1 & 39.39 & 47.44 & 67.44 & 25.30 & 26.80 & 56.98 & 47.46 & 22.61 & 43.52 & 23.08 & \cellcolor{gray!20}\textbf{43.42} & 35.71 \\
    3 &  \cellcolor{gray!20}\textbf{54.55} & 55.13 & 72.09 & \cellcolor{gray!20}\textbf{42.17} & \cellcolor{gray!20}\textbf{34.02} & 70.93 & \cellcolor{gray!20}\textbf{62.71} & \cellcolor{gray!20}\textbf{46.18} & 47.06 & 30.77 & 40.79 & \cellcolor{gray!20}\textbf{49.18}\\
    5 & 42.42 & 55.13 & \cellcolor{gray!20}\textbf{79.07} & 37.35 & 28.87 & \cellcolor{gray!20}\textbf{80.23} & 59.32 & 43.95 & \cellcolor{gray!20}\textbf{50.59} & \cellcolor{gray!20}\textbf{34.62} & 42.11 & 48.57 \\ 
    10 & 36.36 & \cellcolor{gray!20}\textbf{57.69} & 74.42 & 34.94 & 26.80 & 76.74 & 54.23 & 35.35 &  48.24 &  \cellcolor{gray!20}\textbf{34.62} & 40.79 & 44.29\\
    All & 30.30 & 41.03 & 53.49 & 18.07 & 24.74 & 76.74 & 38.98 & 20.06 & 40.00 & 19.23 & 38.16 & 33.06 \\
    \bottomrule
    \end{tabular}
    
    \label{table:size-pass@1}
\end{table*}

%% file: tables/12_strategies_full_repository_pass1.tex
\begin{table*}[!t]
  \centering
  \scriptsize
  \setlength{\tabcolsep}{9pt}
  \renewcommand{\arraystretch}{0.9}
  \caption{Pass@1 and test coverage of \ours (DeepSeek-V3) under different selection strategies on \repocod.}
  \begin{tabular}{@{}l|cccccccccccc|c}
    \toprule
    \multirow{2}{*}{\textbf{Strategies}} 
    & \multicolumn{12}{c|}{\textbf{Pass@1}} 
    & \multicolumn{1}{c}{\textbf{Avg.}}\\
    & \textbf{Sea.} 
    & \textbf{Fla.} 
    & \textbf{Xar.} 
    & \textbf{Sph.} 
    & \textbf{Sym.} 
    & \textbf{More.} 
    & \textbf{Data.} 
    & \textbf{Skl.} 
    & \textbf{Ast.} 
    & \textbf{Pyl.} 
    & \textbf{Plot.} 
    & \textbf{Total} 
    & \textbf{Cov (\%)}\\
    \midrule
    RS  & 44.87 
        & \cellcolor{gray!20}\textbf{76.74} 
        & 32.53 & 42.42 & 28.87 & 68.60 & 44.07 & 12.10 
        & 34.12 & 19.23 & 36.84 & 32.86 & 70.94\\
        
    SS  & 51.28 & 69.77 & 36.14 & 45.45 & 32.99 & 68.60 
        & 47.46 & 15.97 & 36.47 & 23.08 
        & \cellcolor{gray!20}\textbf{43.42} 
        & 36.12 & 72.72\\
        
    FRS & 47.44 & 74.72 & 30.12 & 36.36 & 32.99 & 63.95 
        & 50.84 & 28.98 & 36.47 & 19.23 & 40.79 & 38.88 & 71.58\\
        
    IPS & \cellcolor{gray!20}\textbf{55.13} 
        & 74.42 & 33.73 & 48.48 & 29.90 & 69.77 
        & 50.84 & 28.66 
        & \cellcolor{gray!20}\textbf{47.06} 
        & 26.92 & 42.11 & 41.53 & 76.98\\
    \midrule
    THM & \cellcolor{gray!20}\textbf{55.13} 
        & 72.09 
        & \cellcolor{gray!20}\textbf{42.17} 
        & \cellcolor{gray!20}\textbf{54.55} 
        & \cellcolor{gray!20}\textbf{34.02}
        & \cellcolor{gray!20}\textbf{70.93} 
        & \cellcolor{gray!20}\textbf{62.71} 
        & \cellcolor{gray!20}\textbf{46.18} 
        & \cellcolor{gray!20}\textbf{47.06}
        & \cellcolor{gray!20}\textbf{30.77} 
        & 40.79 
        & \cellcolor{gray!20}\textbf{49.18} 
        & \cellcolor{gray!20}\textbf{79.38}\\
    \bottomrule
  \end{tabular}
  \label{table:strategy-complete-pass@1}
\end{table*}

%% file: sections/evaluation/RQ3.tex
\subsection{RQ3: The Impact of Test Suite Size}
\label{sec:RQ3}

While incorporating more tests may offer richer feedback, it can also introduce redundant signals and increase overhead. To study this trade-off, we investigate how varying the number of selected tests influences the performance of \ours. As a tunable parameter, we set $T=3$ based on preliminary results on the sphinx project (33 tasks) in \repocod. We use the default settings of \ours and only change the number of selected tests $T = \{1, 3, 5, 10, All\}$. Table~\ref{table:size-pass@1} reports Pass@1 across different test suite sizes on \repocod\footnote{Repository abbreviations: Sph.=sphinx, Sea.=seaborn, Fla.=flask, Xar.=xarray, Sym.=sympy, More.=more-itertools, Data.=datasets, Skl.=scikit-learn, Ast.=astropy, Pyl.=pylint, Plot=plotly.py.}. Overall, $T=3$ delivers the highest overall Pass@1 (49.18\%). $T=5$ is optimal for four projects and ranks second overall. Moreover, performance generally declines as the test suite size increases. These results show that our finding on sphinx generalizes to the full benchmark. \textbf{Importantly, providing more tests does not necessarily improve outcomes, and a moderate number of tests provides consistent gains under the TDD setting.}

%% file: sections/evaluation/RQ4.tex
\subsection{RQ4: The Impact of Test Selection Strategies}
\label{sec:RQ4}

In the TDD setting, developers can design tests along multiple dimensions, such as test complexity, whether a test targets fine-grained functionality or spans multiple interacting components, and the form of validation such as assertions. These design choices may lead to tests with different features and feedback signals. Understanding such differences is important not only for agent design but also for informing how developers should construct tests in practice. To this end, we formulate RQ4 to compare THM with four test selection strategies that emphasize different test properties, and analyze how these strategies influence the generation performance.
\subsubsection{\textbf{Random Selection (RS)}}
RS is used as a baseline that uniformly samples a fixed number $T$ of test cases from the full test suite, enabling a direct comparison with more informed selection strategies.

\subsubsection{\textbf{Simplicity-Based Selection (SS)}}
SS prioritizes test cases with low cyclomatic complexity, based on the intuition that simpler tests introduce fewer dependencies and provide clearer failure signals. We compute the cyclomatic complexity of each test function, rank them in ascending order, and select the top-$T$ tests. In \repocod, 66.17\% of tests per task have complexity below 4, indicating that most tests are relatively simple and supporting the practicality of using low-complexity tests to guide LLM-based code generation.

\subsubsection{\textbf{Failure-Revealing Selection (FRS)}}
FRS prefers tests that contain explicit assertions or exception checks. The intuition behind FRS is that tests with explicit validation logic are more likely to expose intended behaviors and produce actionable feedback when failures occur, compared to tests that simply execute code without verification, which may pass silently or fail with ambiguous signals. To implement FRS, we perform AST analysis to identify tests containing assertion-related constructs, including Python \code{assert}/\code{raise}, \code{pytest.raises}, \code{unittest} assertions, and their variants. We then randomly sample $T$ tests from this subset to provide to the agent. In \repocod, 89\% of tests contain assertion-related constructs, indicating that assertion-based tests are prevalent in practice. However, this also enlarges the candidate pool, making FRS behavior closer to RS and potentially reducing its distinctiveness.

\subsubsection{\textbf{Invocation-Proximity Selection (IPS)}}
IPS prioritizes test cases that invoke the target function through shorter call chains. For implementation, we measure the invocation depth of each test case following the THM procedure. We then rank tests in ascending order of depth and select the top-$T$ tests.

Table~\ref{table:strategy-complete-pass@1} reports the complete Pass@1 and average test coverage of \ours{} on \repocod under different test selection strategies when $T=3$. The average test coverage is computed based on the ground-truth target functions. All test selection strategies outperform the RS baseline, with improvements in Pass@1 generally aligning with higher coverage. Among them, \ours{'s} default THM strategy, which emphasizes both invocation proximity and caller diversity, achieves the best results (49.18\% Pass@1 and 79.38\% coverage). This advantage can be attributed to the complementary roles of these two factors. Invocation proximity increases the likelihood that selected tests are semantically relevant to the target function, reducing the reasoning burden to relate the test behaviors with the target implementations. Meanwhile, caller diversity exposes the agent to multiple invocation contexts, enabling it to capture different usage patterns and interaction scenarios. 

\begin{figure}[!t]
    \centering
    \includegraphics[width=\columnwidth,  trim = {0mm 0mm 0mm 0mm}, clip]{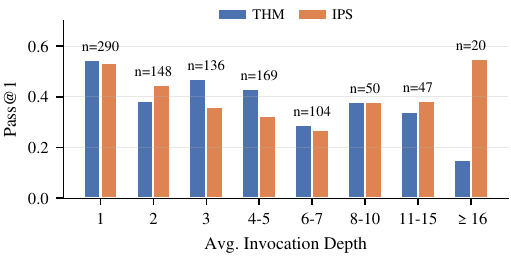}
    \caption{Pass@1 performance of THM and IPS across tasks grouped by average invocation depth in the full test suite.}
    \label{fig:depth-pass}
\end{figure}

To further examine the trade-off between invocation proximity and caller diversity, Fig.~\ref{fig:depth-pass} presents the Pass@1 performance of THM and IPS across tasks grouped by their average invocation depth in the full test suite.
When the average invocation depth is relatively small ($\sim$1–10, covering 913 tasks, 93.16\% of the \repocod), THM consistently outperforms IPS. In this regime, test cases provide sufficiently direct interactions with the target function, allowing THM to effectively leverage diverse invocation contexts and expose \ours to richer usage patterns while maintaining manageable reasoning complexity, resulting in higher generation accuracy.
As the invocation depth increases ($\geq$ 11, 67 tasks, 6.84\% of the \repocod), the advantage of THM diminishes. This suggests that when test suites lack close-range invocations of the target function, prioritizing caller diversity may select tests with long and indirect call chains. Such cases may require the agent to traverse deeper execution paths to interpret test behavior, introducing additional abstractions, noisy signals, and increasing reasoning difficulty.

\textbf{Overall}, THM is effective in the vast majority of tasks with moderate invocation depth, where combining caller diversity and invocation proximity provides strong and complementary guidance for code generation. In scenarios with highly indirect invocation test structures, prioritizing shorter call chains may yield cleaner signals and suggest a promising direction.

%% file: sections/evaluation/RQ6.tex
\subsection{RQ5: The Impact of Test Usage Stage}
\label{sec:RQ5}

Tests can be incorporated at different stages of the agent workflow, potentially influencing generation quality and refinement behavior. To better understand when and how tests contribute, we analyze the impact of using tests at different phases in \ours. In the \textbf{NoTest} setting, no tests are provided and the RRW is disabled. In \textbf{PreGen}, tests are used only during retrieval before the initial generation. In \textbf{PostGen}, tests are applied only in the RRW after the initial generation. Finally, in \textbf{AllStage}, tests are available throughout the workflow. 

\input{tables/9_stage_compare}

\begin{figure}[t]
    \centering
    \includegraphics[width=\columnwidth,  trim = {0mm 0mm 0mm 0mm}, clip]{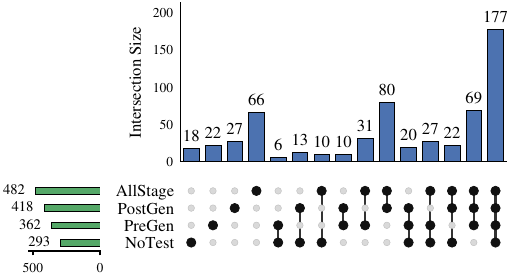}
    \caption{Overlap of solved tasks on \repocod across different test usage stages.}
    \label{fig:overlap}
\end{figure}

Table~\ref{table:stage_sum} reports the performance of leveraging tests at different stages of \ours. We make three main observations. 
\textbf{First, leveraging tests across more stages consistently improves the implementation accuracy.} Pass@1 increases from 29.90\% (NoTest) to 36.93\% (PreGen), 42.65\% (PostGen), and peaks at 49.18\% with AllStage. AllStage performs best on 10 projects, with \code{plotly.py} as the only exception where \textbf{PostGen} is superior. These results demonstrate the effectiveness of incorporating tests throughout the generation process and highlight the potential of TDD for LLM-based software development.
\textbf{Second, this improvement comes at a higher cost}. PostGen and AllStage incur substantially higher computational costs than NoTest and PreGen, requiring nearly 100k additional input tokens, roughly twice the number of output tokens, and more API calls, which reflects the extra context retrieval and debugging efforts required by the RRW.

\textbf{Third, test usage at different stages exhibits complementary effects rather than solving an identical set of tasks}. Fig.~\ref{fig:overlap} shows that AllStage achieves the largest number of uniquely solved tasks (66), while a shared core (177) is solved by all settings. Although part of this diversity may arise from the inherent randomness of LLMs, the substantial partial overlaps (e.g., 80 and 69) indicate that different stages provide complementary rather than redundant signals. Notably, PostGen shares larger intersections with AllStage, suggesting that refinement-time test usage contributes more directly to final correctness. It also points to potential opportunities for improvement by tailoring how tests are leveraged at different stages for different repositories.

%% file: tables/9_stage_compare.tex
\begin{table}[t]
    \centering
    \small
    \setlength{\tabcolsep}{1pt}
    \caption{Results of leveraging tests at different phases in \ours{}'s workflow.}
    \begin{tabular}{lcccc}
        \toprule
        \textbf{Phases} 
          & \textbf{Pass@1 (\%)$\uparrow$}
          & \textbf{Input Cons.$\downarrow$}
          & \textbf{Output Cons.$\downarrow$}
          & \textbf{API Calls$\downarrow$} \\
        \midrule
        NoTest   & 29.90 & 32,829  & 2,482 & 5.04 \\
        PreGen   & 36.93 & 35,427  & 2,408 & 5.63 \\
        PostGen  & 42.65 & 138,330 & 4,710 & 8.91 \\
        AllStage & 49.18 & 147,968 & 4,645 & 8.28 \\
        \bottomrule
    \end{tabular}
    \label{table:stage_sum}
\end{table}

%% file: sections/conclusion.tex
\section{Threats to Validity}
\textbf{Generalizability.}
This work focuses on repository-level code generation under the TDD setting with existing developer-written tests; our findings may not generalize to other settings, such as issue-resolution benchmarks or repositories with incomplete test suites.
\textbf{Test Information Leakage.}
Because existing repository tests are used, evaluation tests may be visible before generation, creating a potential leakage threat: agents may inspect, retrieve, or execute tests that are also used for evaluation. This issue is not unique to \ours{} but is inherited in repository-level code generation benchmarks where evaluation tests remain in the repository during agent execution~\citep{Benchmark_RepoCod, repocoder_repoeval, DevEval}. In \ours{}, we reduce this threat by exposing only THM-selected tests to the downstream agent; all remaining evaluation tests are withheld and used solely for final evaluation.
The THM-selected tests may still leak implementation logic if they reveal the target algorithm rather than observable behavior. We manually examined 20 randomly sampled tasks per benchmark and found no such cases: selected tests mainly specify I/O formats, exceptions, relational properties, or concrete outputs. For example, Fig.~\ref{fig:THM-example} validates RGB outputs while the core RGB$\to$HLS$\to$RGB transformation remains hidden. Still, tests written after implementation may implicitly encode design decisions; a fuller AI-driven TDD workflow should author tests \emph{before} the implementation exists.
\textbf{Evaluation Confounds.}
Agentic systems differ in control flow, tool use, and configuration. We reduce these confounds by aligning tool-calling capacity, monetary budgets, model temperature, and backbone models, but architectural differences and LLM stochasticity remain; repeated runs would further strengthen robustness.
\textbf{Metric Limitations.}
We define correctness as passing the available tests. Since test suites may be incomplete or flaky, our results should be read as test satisfaction under existing developer-written tests rather than full semantic correctness. In RQ4, THM selects tests using heuristic proxies such as cyclomatic complexity and invocation depth, which may not fully capture test quality or fault-revealing capability.

\section{Conclusion \& Future Work}
This work introduces \ours, an agentic framework for \textbf{repository-level code generation under the TDD paradigm}, comprising (1) a test harness mechanism for selecting concise, effective tests, (2) a tailored toolset for context retrieval and debugging, and (3) a reflection-based refinement workflow for iterative code fixing. 
\ours achieves 69.08\% and 81.77\% Pass@1 on \repocod and RepoEval  respectively with Claude Sonnet 4, outperforming baselines by 9.49 and 2.17 percentage points. 
We also present the first systematic study of test suite usage in this setting, analyzing test quantity, selection strategies, and usage stages to inform how TDD can be leveraged in agent-based development. 
Future work includes extending \ours{} from using existing developer-written tests to actively co-evolving tests with implementations, enabling a fuller realization of test-driven repository-level code generation.